\documentclass[prl,twocolumn,showpacs,aps,10pt,superscriptaddress]{revtex4-1} 
\usepackage{graphicx}
\usepackage{amsmath,amssymb,bm} 
\usepackage{epstopdf}
\usepackage{booktabs}
\usepackage{xspace,soul}
\usepackage[usenames,dvipsnames]{color}
\usepackage{subfigure}

\newcommand{\euform}[0]{EuCl\ensuremath{_3}.6H\ensuremath{_2}O\xspace}
\newcommand{\isoeuform}[0]{Eu\ensuremath{^{35}}Cl\ensuremath{_3}.6H\ensuremath{_2}O\xspace}

\newcommand{\eutrans}[0]{{\ensuremath{^7}F\ensuremath{_0\rightarrow ^5}D\ensuremath{_0}}\xspace}
\newcommand{\iso}[1]{\ensuremath{{^{#1}}}} 
\newcommand{\tplus}[0]{\ensuremath{^{3+}}\xspace}

\begin{document}

\title{Ultra-narrow optical inhomogeneous linewidth in a stoichiometric rare earth crystal}

\author{R. L. Ahlefeldt}
\affiliation{Department of Physics, Montana State University, Bozeman, MT 59717, USA}
\affiliation{Laser Physics Centre, Research School of Physics and Engineering, The Australian National University, Canberra 0200, Australia}
\author{M. R. Hush}
\affiliation{School of Engineering and Information Technology,
University of New South Wales at the Australian Defence Force Academy, Canberra 2600, Australia}
\author{M. J. Sellars}
\affiliation{Centre for Quantum Computation and Communication Technology, Research School of Physics and Engineering, The Australian National University, Canberra 0200, Australia}

\date{\today}

\begin{abstract}
We have obtained a low optical inhomogeneous linewidth of 25 MHz in the stoichiometric rare earth crystal \euform by isotopically purifying the crystal in \iso{35}Cl. With this linewidth, an important limit for stoichiometric rare earth crystals is surpassed: the hyperfine structure of \iso{153}Eu is spectrally resolved, allowing the whole population of  \iso{153}Eu\tplus ions to be prepared in the same hyperfine state using hole burning techniques. This material also has a very high optical density and  can have  long coherence times when deuterated. This combination of properties offers new prospects for quantum information applications. We consider two of these, quantum memories and quantum many body studies. We detail the improvements in the performance of current memory protocols possible in these high optical depth crystals, and how certain memory protocols, such as off-resonant Raman memories, can be implemented for the first time in a solid state system. We explain how the strong excitation-induced interactions observed in this material resemble those seen in Rydberg systems, and describe how these interactions can lead to quantum many-body states that could be observed using standard optical spectroscopy techniques.
\end{abstract}


\maketitle
The hyperfine levels of rare earth ions have exceptionally long coherence times, as much as 6 hours in \iso{151}Eu\tplus \cite{zhong15}. Additionally, these hyperfine levels are addressable through an intermediary optical level, allowing high fidelity spin storage and readout using optical pulses \cite{jobez15,gundogan15}. Because of these properties, doped rare earth crystals have received considerable attention for quantum memory applications \cite{gundogan15, jobez15, hedges10, dajczgewand14, saglamyurek15}.

All the spin-wave quantum memory demonstrations to date have used rare-earth doped crystals with low concentrations, ranging from 0.001 to 0.05\% \cite{hedges10, afzelius10, gundogan15, jobez15}, resulting in low optical densities of the order of 1~cm$^{-1}$. In addition, partly as a consequence of the disorder introduced by the rare earth dopant, these crystals exhibited inhomogeneous linewidths of the order of GHz, much greater than the hyperfine structure and the available Rabi frequencies. As a result, spectral hole burning techniques were necessary to create the narrow spectral features that would allow  memory demonstrations. The use of these techniques, combined with the low initial optical density of the crystal, severely reduces the number of ions that can contribute to the operation of the memory (the usable optical depth), limiting its efficiency and storage density \cite{gorshkov07universal, nunn08, jobez16}. Because of these limitations, only two quantum memory demonstrations to date have surpassed the no-cloning limit, and in both cases a very large interaction length was required to obtain a sufficient optical depth to store even a single mode \cite{hedges10,schraft16}.

Narrow linewidth, high optical density materials would therefore be very beneficial for quantum memory applications. Very narrow inhomogeneous linewidths have been observed in rare-earth doped crystals by reducing the dopant concentration to the ppm level and by isotopically purifying the host.  The most well-known example is YLiF$_4$ \cite{agladze91, macfarlane92, chukalina00}. When isotopically purified in $^7$Li to remove broadening caused by isotopic disorder, this material has optical inhomogeneous linewidths as low as 16~MHz for Er\tplus \cite{thiel11rare} and 10~MHz for Nd\tplus \cite{macfarlane98}. To achieve such narrow lines, it was necessary to use extremely low rare earth dopant concentrations, around 3~ppm. So although these crystals are much lower disorder than current memory materials, their optical and spatial densities are low, and little advantage is gained from their narrow lines for quantum memory purposes.

In this paper, we show that small inhomogeneous linewidths can be achieved at the same time as high optical and spatial densities by isotopically purifying a material that is stoichiometric, rather than doped, in the rare earth ion. The linewidth reduction is sufficient in our chosen material, \euform, that the rare earth hyperfine structure is resolved. Reaching this limit in the high optical depth regime has broad-reaching implications for quantum memories. It also opens new applications, and in particular we discuss the use of these materials for many body physics studies, which are made possible because the linewidth is also smaller than the nearest-neighbor ion-ion interactions \cite{ahlefeldt13precision}.

\euform is a good starting point in the quest to achieve ultra-narrow linewidths because it already has a very narrow, 100 MHz optical linewidth without isotopic purification\cite{ahlefeldt09}, and good coherence times when deuterated \cite{ahlefeldt13optical}. This makes it an excellent quantum memory candidate in its own right.

\begin{figure}
\centering
\includegraphics[width = 0.4\textwidth]{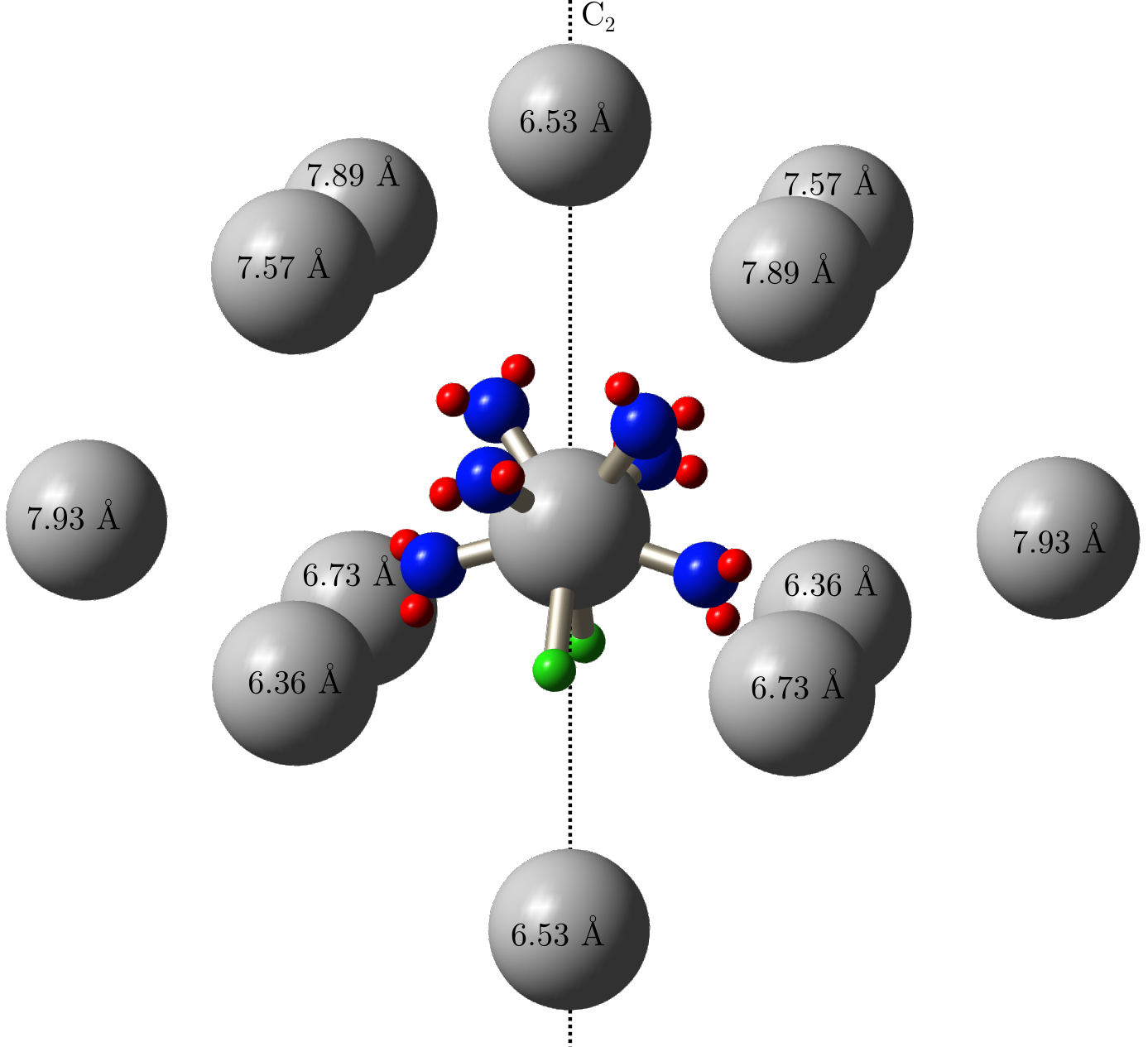}
\caption{\label{fig:eucl3structure} The first shell of Eu\tplus ions (gray) in \euform. The C$_2$ symmetry axis is vertical. The Eu\tplus ions are labeled by their distance from the central ion. Ligand ions Cl (green), O (blue) and H (red) are only shown for the central ion. Our previous measurements suggest that the static interaction between ions separated by 6.53~\AA\xspace along the C$_2$ axis is $>40$~MHz.   }
\label{figEucystal}
\end{figure}

\euform is a monoclinic crystal with  a single rare earth site of C$_2$ site symmetry \cite{belskii65, kepert83}. The environment of an Eu\tplus ion in \euform is shown in Figure \ref{figEucystal}. It has eight direct ligands -- six water molecules and two Cl ions at distances between 2.4 and 2.8 \AA. Also shown are the first shell of Eu\tplus neighbors.

The \eutrans transition of \euform at 579.703~nm (vacuum) shows a  structured line spread over 600 MHz, with each component line approximately 100 MHz wide. The structure of the line can be completely explained by a combination of the hyperfine structure of the two europium isotopes and shifts caused by different isotopes occupying the nearest neighbor ligand positions \cite{ahlefeldt09}. These isotope shifts can be large, up to 2~GHz for Eu\tplus ions neighboring D ions (\iso{2}H), and we expect that isotopes occupying more distant sites contribute substantially to the inhomogeneous broadening.

The contribution of the different isotopes to the broadening can be estimated simply. It  depends on the concentration of that element in the crystal, the abundance of the isotope relative to the dominant isotope, and the perturbation to the lattice caused by substituting the isotope. This latter quantity depends on the relative mass difference of the isotope and the dominant isotope, thus D causes the largest perturbation and \iso{151}Eu\tplus the smallest.

We have measured the inhomogeneous broadening rate due to D as 91 MHz/\% concentration \cite{ahlefeldt09}, suggesting that D is responsible for 1.4~MHz of the inhomogeneous broadening seen in natural abundance crystals. Assuming the broadening scales linearly with the relative mass difference, the concentration and the abundance, we expect that \iso{17}O and \iso{18}O contribute $\approx 2$~MHz to the inhomogeneous broadening, \iso{37}Cl $\approx 60$~MHz, and \iso{151}Eu\tplus $\approx 10$~MHz. These numbers are approximate, but clearly indicate that \iso{37}Cl is likely to be the major source of broadening in \euform, and that the best method for reducing the linewidth is to isotopically purify the crystal in \iso{35}Cl.

We grew a crystal isotopically purified in \iso{35}Cl from a water solution at just above room temperature. The starting material was 6 g of \isoeuform that had been prepared from Na\iso{35}Cl and 99.999\% \euform. The nominal isotopic purity was 99.67\%. Sufficient water was added to saturate the solution at 30$^\circ$C. A crystal was grown out of the solution over 1 week after lowering the temperature to 29$^\circ$C. The growth rate during this time was not constant as the hotplate used for growth only has a temperature stability of $\pm 1^\circ$C. An unstable growth rate can lead to inclusions and other growth defects. Because the base linewidth in this material is so low, the effect of these rare growth defects can start to be seen as small macroscopic variations in the inhomogeneous linewidth of the order of 20 MHz over millimeter-sized regions of the crystal. We expect that the majority of these defects and the associated macroscopic variations in broadening can be removed by improving the temperature stability of the growth process.

 The excitation spectrum of the \eutrans transition was measured with a Coherent 699-29 ring dye laser with the crystal at 4~K in a helium bath cryostat. Because \euform is hygroscopic, the crystal was mounted in a small helium-gas-filled chamber to avoid exposure to air or vacuum. The laser beam was orientated parallel to the C$_2$ axis to minimize absorption. Even along this direction, the absorption is high, and a confocal imaging system  was used to collect fluorescence only from the front face of the crystal. In this setup, achromatic 10~cm lenses were used for objective and imaging lenses, and the emission was focused onto the end of a 62.5~$\mu$m multimode fiber serving as the pinhole.

\begin{figure}
\centering
\includegraphics{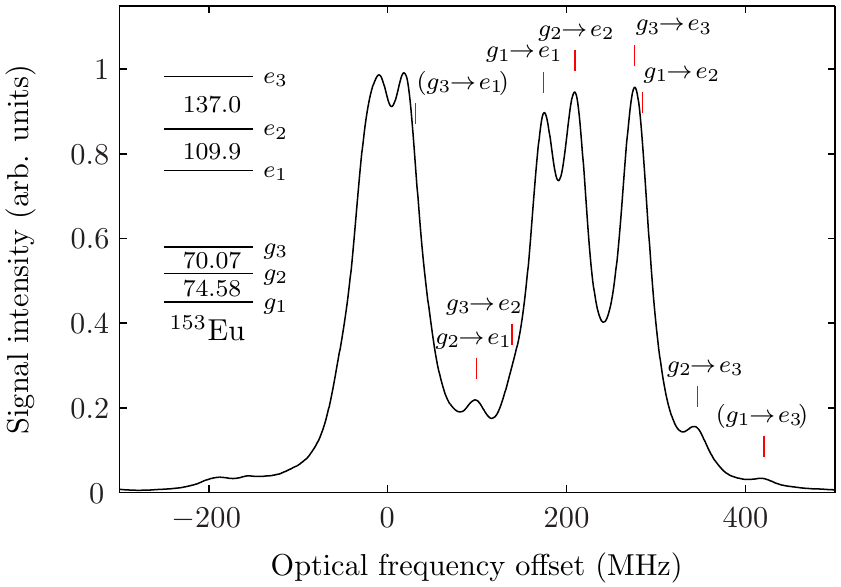}
\caption{\label{fig:iso}  Excitation spectrum of the \eutrans transition in isotopically pure \isoeuform. The energy level structure of the $^7$F$_0$ ground and $^5$D$_0$ excited states for \iso{153}Eu\tplus is shown on the upper left, where the splittings are given in MHz. The positions of each transition between the \iso{153}Eu\tplus ground and excited states are indicated by red lines on the spectrum. The weak transitions are bracketed. }
 \end{figure}

 Figure \ref{fig:iso} shows the excitation spectrum of \isoeuform. There is a 200 MHz Eu\tplus isotope shift in \euform, leading to two sets of peaks: around 0 MHz frequency offset are those due to \iso{151}Eu\tplus, while at higher frequencies are those due to \iso{153}Eu\tplus. These sets of peaks are split by the hyperfine interaction. For \iso{153}Eu\tplus, which has hyperfine splittings approximately twice as large as \iso{151}Eu\tplus, this hyperfine structure is well resolved. In Fig. \ref{fig:iso}, we have labeled each peak for \iso{153}Eu\tplus with the transition that generates it.
 
 The inhomogeneous linewidth of each line in the spectrum is $25\pm 5$~MHz. The instrumental broadening due to instability in the laser scan is less than 5~MHz. As described earlier, we expect that about 10~MHz of the remaining linewidth is due to \iso{151}Eu\tplus, with about 3~MHz due to the extra O and H isotopes. Therefore, isotopic purification of the other elements in the crystal will reduce the linewidth to the order of 10~MHz. Further purification also has the added advantage that it can be used to tune some of the crystal properties. For example, fully deuterating the crystal will increase the coherence times \cite{ahlefeldt13optical}, while altering the 50:50 europium isotope ratio can be used to tune the Eu\tplus ion-ion interactions.
 
At 25~MHz, the inhomogeneous broadening has already reached an interesting limit: it is smaller than the \iso{153}Eu\tplus hyperfine structure, as well as the static excitation-induced interactions between Eu\tplus ions. This regime is very useful for certain quantum information applications. We will discuss two of these: quantum memories and quantum many-body dynamics. 


There are two properties of \isoeuform that are important for both these applications. First, long coherence times on the optical and hyperfine transitions. An optical coherence time of 700~$\mu$s  \cite{ahlefeldt13optical} is obtainable with deuteration, while the zero first order Zeeman (ZEFOZ) technique can be used to get very long hyperfine coherence times. Second, the ability to optically pump the majority of \iso{153}Eu\tplus ions into a single hyperfine state, which is possible because of the narrow linewidth. This results in a higher optical depth, and initializes the ions in a well-defined state, which is the capability that enables many body investigations and certain memory protocols in this system. To estimate the proportion of the population that can be pumped into a single hyperfine state, we use a simple rate equations model in which the crystal is excited by two fields: an optical field swept over the $g_1\rightarrow e_3$ and $g_2\rightarrow e_3$ transitions and an rf field on the $g_1\rightarrow g_2$ transition. Assuming a very long hyperfine lifetime, a valid assumption for \euform which has a lifetime of several hours, over 90\% of the \iso{153}Eu\tplus ions in the entire crystal can be pumped into $g_3$, and at the center of the $g_3\rightarrow e_3$ line over 98\% of the ions are in $g_3$. Given the oscillator strength that we have previously measured for \euform ($3\times10^{-9}$), the peak optical density of the prepared feature is $>4000$~cm$^{-1}$.

Combined with \isoeuform's other properties, this extremely high optical density makes this crystal very attractive for quantum memory applications. The low optical density of current memory crystals, of the order of 1~cm$^{-1}$, has severely limited the quantum memory performance in demonstrations to date. An increase in optical density can improve aspects of the performance of all spin-wave storage quantum memory protocols for rare earth crystals, which comprise  the atomic frequency comb (AFC) \cite{afzelius09}, the gradient echo memory (GEM) \cite{alexander07, hetet08} and Raman-type protocols such as electromagnetically induced transparency (EIT) \cite{fleischhauer02,harris90}. Here, we briefly describe the improvements in memory performance that can be expected from the high optical density and other properties of \isoeuform.

The quantum memory efficiency for all the protocols above is directly dependent on the optical depth \cite{hush13,nunn08}. The efficiency in memory demonstrations to date has been limited by the low optical density to less than 80\% \cite{hedges10, schraft16}, with most demonstrations being well below the no-cloning limit, 50\%. The high optical density available in \isoeuform means a memory efficiency of above 90\% should be easily achievable, exceeding the highest memory efficiency seen, 87\% in Rb gas \cite{cho16}.

The high optical depth of \isoeuform also allows a much high spatial multiplexing capacity than current materials. \isoeuform will show appreciable absorption over 1~$\mu$m, allowing the creation of a dense set of spatial voxels, each capable of storing a mode at high efficiency.

Efficiency and spatial multiplexing capacity are improved across all protocols. Improvements to other performance metrics are possible depending on the protocol. For example, while the GEM and EIT protocols allow higher initial efficiency that the AFC protocol,  the bandwidth is dependent on the optical depth, so there is eventually a trade-off between efficiency and bandwidth or spectral/temporal multiplexing capacity \cite{nunn08}. This trade-off is alleviated by a very high optical depth, allowing large bandwidth, high efficiency storage. For GEM, a high optical density should allow the bandwidth limit, the ground state hyperfine splitting, to be reached. This limit is generally considered to apply to all spectral-holeburning based memories like GEM and AFC. However, for a material like \isoeuform in which the hyperfine structure is resolved, this limitation is lifted in the GEM protocol \cite{hedges10}, allowing much larger memory bandwidths.

The ability to resolve the hyperfine structure present in \isoeuform also allows new memory protocols to be implemented in this material. In particular, off-resonant Raman schemes \cite{reim10, moiseev13} become possible for the first time in a solid state system. These protocols require a high optical depth, and the ability to completely empty one hyperfine state over the entire spectral width of the inhomogeneous line, which is only possible when the hyperfine structure is completely involved. Off-resonant Raman protocols have a number of advantages over the resonant schemes described above, such as lower noise because the excited state is never populated.

Similar to Raman memories, the second quantum information application we will consider, quantum many body dynamics, is enabled in \isoeuform by the high optical density and small optical inhomogeneous broadening. In this material, the Eu\tplus atoms are close together and can have interactions much stronger than the residual inhomogeneity in the system. The result is a lattice of interacting Eu\tplus atoms, the ideal platform for investigating many-body quantum phenomena. 

The interactions that dominate in this material will determine what many-body states can be observed. We are interested in interactions on the optical transition between the $^7$F$_0$ ground state and the $^5$D$_0$ electronic excited state. We assume the \iso{153}Eu\tplus population has been prepared in the $g_3$ state using the holeburning method described above. Two Eu\tplus ions separated by a distance $r$ can interact through a variety of complex mechanisms including direct electric multipole interactions or exchange, a wavefunction-overlap mechanism that is often mediated by intervening ligands (superexchange). Regardless of the mechanism, there are two qualitatively different parts of the interaction. The first is a diagonal interaction producing a static shift to the electronic levels $H_{\rm d}/h = V(r) |e_3, e_3 \rangle \langle e_3,e_3|\,$. This interaction is the nearest-neighbor version of the interaction that commonly leads to instantaneous spectral diffusion in rare earth crystals \cite{liu87,liu90laser}. We previously measured these nearest neighbor interactions to be on the order of 40~MHz in \euform \cite{ahlefeldt13precision}, larger than the inhomogeneous linewidth. The second interaction is an off-diagonal interaction, of the form $H_{\rm o}/h = T(r) |e_3, g_3 \rangle \langle g_3, e_3 | + T^*(r) |g_3, e_3 \rangle \langle e_3, g_3 |$.

Using our experimental measurements, we can estimate the relative strength of the off-diagonal $T$ and diagonal $V$ interactions in two different regimes: long and short range. At long range, we can assume the dipole-dipole interaction will dominate. Hence, the relative strength of the interactions will be the ratio of the static $d_{\rm trans} = 1.6 \times 10^{-33}$ C.m and transition $d_{\rm static} = 1.0 x 10^{-32}$ C.m dipole moments squared \cite{ahlefeldt13precision}. Thus, the diagonal interaction will be $d_{\rm static}^2/d_{\rm trans}^2 = 42$ times stronger than the off-diagonal interactions. At short ranges, if there were strong off-diagonal interactions we would expect to see the linewidth of the spectrum increase as the ion density increases due to the contribution of these interactions. However, in our experiments we did not see any significant difference in the linewidth of the bulk line  in Fig. \ref{fig:iso}), where the Eu\tplus ion density is high, and the satellite lines due to \iso{18}O (not visible in the figure), where the Eu\tplus density is very low.  Hence, we can estimate the short range off-diagonal interactions are no larger than 5 MHz, the accuracy of our linewidth measurements, which is a factor of 10 smaller than the largest observed diagonal interaction. Therefore, in either regime, we expect the static shift interaction will dominate the many body dynamics observed.

As the diagonal interaction dominates, a close analogue for \isoeuform among cold atom systems is an optical lattice of Rydberg atoms \cite{anderson11,viteau11}, and the many-body effects predicted for Rydberg systems should also be observable in our rare earth ion setting. Currently, the primary experimental signature used to verify the presence of Rydberg interactions is the blockade effect \cite{dudin12,tong04}: exciting one atom in the system pushes nearby atoms out of resonance with the exciting field. Since the ion-ion distances and the associated blockade region are much smaller in solid state systems, the spatial imaging methods used to study Rydberg systems are not applicable, and spectral imaging methods must be used.

The blockade effect could be indirectly observed through the suppression of excitation \cite{tong04} or EIT spectroscopy \cite{pritchard10}. To directly determine the blockade region, a spectrally narrow sub-ensemble of ions can be excited, and then the shifts caused by this excitation can be observed in the optical spectrum \cite{ahlefeldt13precision}. The geometry of the interaction strengths can be determined from the spectrum and used to infer the exact size of the blockade region. Lastly, although the size of the blockade region is likely to be well below the diffraction limit of light, sub-wavelength imaging is possible by using a spectro-spatial method \cite{schiller92},  in which an electric or magnetic field gradient creates a large spatial variation in the optical transition frequency over a region much smaller than the diffraction limit. 

After verification and quantification of the blockade region, there are a variety of many-body states theoretically predicted to exist for Rydberg systems that could be investigated in rare earth crystals. For example, paramagnetic states with short range quantum correlations \cite{lesanovsky11} and the so-called devil's staircase of crystalline phases \cite{weimer10, pohl10, schachenmayer10, pupillo10}. More general many-body phenomena have also been predicted, but are yet to be confirmed, such as quantum critical behavior \cite{weimer08}, fermionic transport \cite{olmos09}, collective jumps \cite{lee12} and non-equilibrium dynamical transitions \cite{marcuzzi14}. 

Recently, anomalous broadening has been observed in lattices of Rydberg atoms prepared with a two-photon transition, which gets rapidly larger with increasing atomic density \cite{goldschmidt16}. This broadening will not be present in the rare earth ion crystal because a conventional single photon transition is used to access the interacting states which are spectrally distinct. This means that these rare earth ion crystals may be able to probe coherent interacting dynamics in the dense atomic regime, which now appears to be unavailable in Rydberg systems. 

Here we have concentrated on the use of \isoeuform for two quantum information applications, quantum memories and quantum many body studies. The low disorder and high optical density of this material could also be useful  for other applications, such as optical delay line protocols like the two level AFC \cite{afzelius09}, and DLCZ-type repeater protocols \cite{duan01,ledingham10}. In addition, it is likely that similar properties are obtainable in other stoichiometric rare earth materials. Having a range of materials is particularly useful for quantum many body studies, as the variation in crystal structure and types of interaction should provide a way to probe different many body effects. 

In conclusion, we have shown that \euform isotopically purified in \iso{35}Cl has an optical inhomogeneous linewidth for the \eutrans transition of Eu\tplus of 25~MHz, the lowest of any stoichiometric rare earth crystal and one of the lowest inhomogeneous linewidths measured in any material. Isotopic purification of the other elements in the crystal, in particular Eu\tplus, will further lower this linewidth. The small linewidth and high optical depth mean this material would be a good system for quantum memories. Combined with these properties, the strong excitation induced frequency shifts between nearest neighbors suggest that many-body physics analogous to that seen in Rydberg systems can be observed.

\begin{acknowledgments}
This work was supported by the Australian Research Council Centre of Excellence for Quantum Computation and Communication Technology (Grant No. CE110001027). MRH acknowledges funding from an Australian Research Council (ARC) Discovery Project (project number DP140101779). MJS was supported by an Australian Research Council Future Fellowship (Grant No. FT110100919). The authors thank E. A. Goldschmidt for helpful discussions.
\end{acknowledgments}


%

\end{document}